\documentclass[dvipsnames]{IEEEtran}

\usepackage{etoolbox}

\renewcommand{\baselinestretch}{1}

 \newif\ifOneColumn
 \OneColumnfalse

\def \mydis {19.1mm}
\def \mydisbottom {19.1mm}
\def \mytopdis {1in}
\def \mytopotherdis {0.75in}

\usepackage[papersize={8.5in,11in},top=\mytopotherdis,left=\mydis,right=\mydis,bottom=\mydisbottom]{geometry}
\renewcommand{\baselinestretch}{1}

\usepackage{etex}
\reserveinserts{28}
\usepackage{makeidx}         
\usepackage{graphicx}        
\usepackage{multicol}        
\usepackage{multirow}
\usepackage[bottom]{footmisc}
\makeindex             

\usepackage{enumitem, fancyhdr, amsthm, amsmath, amsfonts,  amssymb, csquotes, mathtools, url, graphicx, balance, comment, mathrsfs, upgreek, wrapfig, cases, pgfplots,  fancybox,  multirow, hhline, mathtools, chngcntr, booktabs, algpseudocode, mathabx, floatrow, mathrsfs, etex, etoolbox, xfrac,  cite, xstring, xparse, ifmtarg, xifthen, bm, array, multicol, layouts, printlen, soul, cancel, tabularx, ragged2e, accents, caption, nomencl, nccmath}
\makenomenclature
\renewcommand\nomgroup[1]{%
	\item[\bfseries
	\ifstrequal{#1}{A}{DN Parameters}{%
		\ifstrequal{#1}{B}{Parameters of edge $e\in\E$}{%
			\ifstrequal{#1}{C}{Nodal quantities of node $i\in\N$}{%
				\ifstrequal{#1}{D}{Quantities of load at node $i\in\N$}{%
					\ifstrequal{#1}{E}{Quantities of DER $\der\in\setDER$}{%
						\ifstrequal{#1}{F}{Quantities for failure scenario $\scenarioIdx$}{%
							\ifstrequal{#1}{G}{Allocation decision variables}{%
								\ifstrequal{#1}{H}{Repair decision variables for scenario $\scenarioIdx$}{%
									\ifstrequal{#1}{I}{Other Symbols}{%
	}}}}}}}}}%
]}

\usepackage[ruled,vlined]{algorithm2e} 
\usepackage[]{subfig}
\usepackage{array,verbatim,multirow,bbm,float}
\usepackage[]{todonotes}
\usepackage[normalem]{ulem}
\usepackage{tikz}
\usetikzlibrary{decorations.pathreplacing}
\usetikzlibrary{patterns,arrows,shapes,snakes,automata,backgrounds,petri,plotmarks}
\usepackage{color,xcolor}
\usepackage[colorlinks=true,
raiselinks=true,
linkcolor=MidnightBlue,
citecolor=Mahogany,
urlcolor=ForestGreen,
urlcolor=black,
pdftitle={Optimal Resource Allocation in Electricity Distribution Networks During Contingencies},
pdfkeywords={DN Security},
pdfsubject={Initial Manuscript}
]{hyperref}

\usepackage[nameinlink]{cleveref}
\crefname{section}{Sec.}{§§}
\Crefname{section}{Sec.}{§§}
\crefname{subsection}{Sec.}{§§}
\Crefname{subsection}{Sec.}{§§}
\crefname{equation}{Eq.}{}
\Crefname{equation}{Eq.}{Eqs.}
\crefname{appsec}{Appendix}{Appendices}
\crefname{table}{Table}{Tables}
\crefname{figure}{Fig.}{Figures}
\crefname{theorem}{Theorem}{Theorems}
\crefname{proposition}{Proposition}{Propositions}
\crefname{lemma}{Lemma}{Lemmas}
\crefname{algorithm}{Algorithm}{Algorithms}
\crefname{myclaim}{Claim}{Claims}

\def \sspace{-3pt}
\def \ssspace{-3pt}
\usepackage[nodisplayskipstretch]{setspace}

\newcommand{\abs}[1]{\lvert{#1}\rvert}

\allowdisplaybreaks

\makeatletter
\renewcommand*\env@matrix[1][*\c@MaxMatrixCols c]{%
	\hskip -\arraycolsep
	\let\@ifnextchar\new@ifnextchar
	\array{#1}}
\makeatother

\newcommand{\myc}[3]{
	\def \firstString {{#1}}	
	\IfEqCase{#2}{
		{}{\firstString_{#3}}
		{t}{\firstString_{#3}^{t}}
		{T}{\firstString_{#3}^{T}}
		{\top}{\firstString_{#3}^{\top}}
		{\site}{\firstString_{#3}^{\site}}
		{\ts}{\firstString_{#3}^{\ts}}
		{\scenarioIdx}{\firstString_{#3}^{\scenarioIdx}}
		{\ts-1}{\firstString_{#3}^{\ts-1}}
		{\period}{\firstString_{#3}^{\period}}
		{\period-1}{\firstString_{#3}^{\period-1}}
		{\period,\scenarioIdx}{\firstString_{#3}^{\period,\scenarioIdx}}
		{0,\scenarioIdx}{\firstString_{#3}^{0,\scenarioIdx}}
		{\nperiod,\scenarioIdx}{\firstString_{#3}^{\nperiod,\scenarioIdx}}
		{\period-1,\scenarioIdx}{\firstString_{#3}^{\period-1,\scenarioIdx}}
		{0}{\firstString_{#3}^{0}}
		{\opr}{\firstString_{#3}^{\opr}}
		{\con}{\firstString_{#3}^{\con}}
		{\star}{\firstString_{#3}^{\star}}
		{n}{\firstString_{#3}}
		{l}{\widehat{\firstString}_{#3}}
		{u}{\widecheck{\firstString}_{#3}}
		{max}{\mathbf{\overline{\firstString}}_{#3}}
		{min}{\mathbf{\underline{\firstString}}_{#3}}
		{\hardMax}{\bm{\overline{\overline{\firstString}}_{#3}}}
		{\hardMin}{\bm{\underline{\underline{\firstString}}_{#3}}}	
		{constant}{\mathbf{\firstString}_{#3}}
		{\pre}{\firstString_{#3}^{n}}
		{\post}{{\firstString}_{#3}^{c}}
		{act}{\firstString_{#3}^{act}}
		{set}{\firstString_{#3}^{set}}
		{ref}{\firstString_{#3}^{\text{ref}}}
		{r}{\firstString_{#3}^{r}}
		{t-1}{\firstString_{#3}^{t-1}}
		{nom}{\mathbf{\firstString_{#3}^{nom}}}
		{stab}{\bm{\firstString_{#3}^{stab}}}
		{devmax}{\firstString_{#3}^{dev,max}}
		{reg}{\bm{\firstString_{#3}^{reg}}}
		{ev}{\firstString_{#3}^{ev}}
		{nev}{\firstString_{#3}^{nev}}
		{maxev}{\bm{\overline{\firstString}_{#3}^{ev}}}
		{maxnev}{\bm{\overline{\firstString}_{#3}^{nev}}}
		{\state}{\firstString_{#3}^{\state}}
		{\attack}{\firstString_{#3}^{\attack}}
		{\optimalAttack}{\firstString_{#3}^{\attack\star}}
		{\defend}{\firstString_{#3}^{\defend}}
	}[\firstString_{#3}^{#2}]
	{
	}
}

\DeclareMathOperator*{\argmin}{arg\,min}

\def \nperiod {\mathrm{K}}
\def \period {k}
\def \nscenario {\mathrm{S'}}

\def \scenario {s}

\def \scenarioIdx {s}

\def \node {i}
\def \site {i}
\def \j {\mathbf{j}}
\def \ts {\period,\scenarioIdx}

\def \setSite {\mathcal{U}}
\def \setScenarios {\mathcal{S}}
\def \setScenariosSmall {\setScenarios'}

\newcommand{\genBudget}{\mathrm{G}{}{}}
\newcommand{\crewBudget}{\mathrm{Y}{}{}}
\def \setPeriods {\mathcal{K}}
\def \setBinary {\mathbb{B}}

\def \first {a}
\def \First {\mathcal{A}}

\newcommand{\soo}[2]{g_{#1}^{#2}}

\newcommand{\costFirstStageCoeff}{c_\text{a}}
\newcommand{\costLoad}[1]{\text{L}^{#1}}

\def\ma  {{\text{m}_\text{a}}}
\def\na  {{\text{n}_\text{a}}}
\def\nx {{\text{n}_\text{x}}}

\def\nys {{\text{n}_\text{y}^s}}

\newcommand{\costFirstStage}{\text{J}^\text{I}}
\newcommand{\costSecondStage}{\text{J}^\text{II}}
\newcommand{\costSecondStagePeriod}[1]{\text{J}^{#1}}

\newcommand{\valueSecondStagePeriod}[1]{\text{V}^{#1}}
\newcommand{\probFails}{\myc{\mathcal{P}}{}{}}

\newcommand{\recourseUb}{\Phi}

\newcommand{\costSecondStageLin}[2]{\myc{c}{#1}{\text{#2}}}

\newcommand{\firstStageConstraints}{\text{A}_\text{a}}
\newcommand{\firstStageRH}{b}
\newcommand{\secondStageConstraints}[2]{\myc{\text{B}}{#1}{\text{#2}}}

\newcommand{\secondStageRh}{\myc{\mathrm{h}}{}{}}
\newcommand{\secondStageRhA}{\myc{\mathrm{T}}{}{}}

\newcommand{\indexSet}[1]{\mathcal{I}_{\text{#1}}}

\def \systemPerformance {\mathcal{R}^{\period}}

\def \disc {\mathrm{d}}
\def \cont {\mathrm{c}}

\newcommand{\lossLP}{\mathrm{L^\star}}

\newcommand{\saaSoo}[2]{\hat{g}_{#1}^{#2}}

\newcommand{\iter}{r}

\newcommand{\niter}{R}

\newcommand{\bruteforce}{\mathrm{SE}}
\newcommand{\bwga}{\mathrm{LBDwGA}}
\newcommand{\bora}{\mathrm{BoRA}}
\newcommand{\sa}{\mathrm{SA}}
\newcommand{\se}{\mathrm{SE}}

\def \setDER {\mathcal{D}}
\def \der {d}
\def \ref {\text{ref}}

\newcommand{\N}[1][]{\mathcal{N}_{#1}}
\newcommand{\E}[1][]{\mathcal{E}_{#1}}
\newcommand{\G}[1][]{\mathcal{G}_{#1}}

\newcommand{\fromNode}[1]{{#1}^{-}}
\newcommand{\toNode}[1]{{#1}^{+}}

\def \bigM {\mathrm{L}}

\def \R {\mathbb{R}}

\def \x {{x}}

\def \state {\eta}
\def \opr {o}
\def \con {c}
\def \hardMin {hmin}
\def \hardMax {hmax}

\def \NN {{\mathrm{N}}}

\def \C {{\text{W}}}

\def \Csite {\C^\text{\footnotesize SD}}
\def \Cshed {\C^\text{\footnotesize LS}}
\def \Cload {\C^\text{\footnotesize LC}}

\def \Xnpf {\mathcal{X}}

\newcommand{\resistance}[1]{\text{r}_{#1}} 
\newcommand{\reactance}[1]{\text{x}_{#1}}

\def \ssum {\textstyle\sum}

\def \mmin {\textstyle\min}

\usepackage{chngcntr}
\counterwithout{subsubsection}{subsection}

\makeatletter
\@addtoreset{subsubsection}{subsection}
\makeatother

\usepackage{tcolorbox}

\usepackage{mdframed}

\newcommand{\nuc}[2]{\myc{\mathrm{v}}{#1}{#2}}

\def \pre {pre}
\def \post {post}

\newcommand{\pfc}[2]{\myc{\eta}{#1}{#2}}	

\newcommand{\kcc}[2]{\myc{kc}{#1}{#2}}	
\newcommand{\kgc}[2]{\myc{y}{#1}{#2}}	
\newcommand{\klinec}[2]{\myc{kl}{#1}{#2}}

\newcommand{\Pc}[2]{\myc{P}{#1}{#2}}
\newcommand{\Qc}[2]{\myc{Q}{#1}{#2}}

\newcommand{\nucc}[2]{\myc{vc}{#1}{#2}}
\newcommand{\nugc}[2]{\myc{vg}{#1}{#2}}

\newcommand{\pcc}[2]{\myc{pc}{#1}{#2}}
\newcommand{\qcc}[2]{\myc{qc}{#1}{#2}}

\newcommand{\ptc}[2]{\myc{p}{#1}{#2}}
\newcommand{\qtc}[2]{\myc{q}{#1}{#2}}

\newcommand{\kqc}[1]{\myc{mq}{constant}{#1}}

\newcommand{\ysc}[2]{\myc{u}{#1}{#2}}
\newcommand{\ygc}[2]{\myc{w}{#1}{#2}}
\newcommand{\ylinec}[2]{\myc{yl}{#1}{#2}}

\newcommand{\pgc}[2]{\myc{pg}{#1}{#2}}
\newcommand{\qgc}[2]{\myc{qg}{#1}{#2}}

\newcommand{\xc}[2]{\myc{\x}{#1}{#2}}
\newcommand{\yc}[2]{\myc{y}{#1}{#2}}

\newcommand{\Xc}[2]{\myc{\Xnpf}{#1}{#2}}
\newcommand{\Yc}[2]{\myc{\mathcal{Y}}{#1}{#2}}
\newcommand{\lcc}[2]{\myc{\beta}{#1}{#2}}

\newcommand{\transpose}[1]{{#1}^{\top}}

\newcommand{\mycc}[4]{
	\def \firstString {{#1}}	
	\IfEqCase{#2}{
		{n}{\firstString_{#3}^{#4}}
		{max}{\overline{\firstString}_{#3}^{#4}}
		{min}{\underline{\firstString}_{#3}^{#4}}
	}
}

\ifOneColumn
	\renewcommand{\baselinestretch}{2}
\else

	\renewcommand{\baselinestretch}{0.94}
	\def \sspace{-1pt}
	\def \ssspace{-1pt}
	\newcommand{\subparagraph}{}
	\usepackage[]{titlesec}
	\titlespacing{\section}{0pt}{-\sspace}{-\ssspace}
	\titlespacing{\subsection}{0pt}{-\sspace}{-\ssspace}
	\titlespacing{\subsubsection}{0pt}{-\sspace}{-\ssspace}
	
\titleformat{\paragraph}[runin]{\normalfont\normalsize\itshape}{\theparagraph}{1em}{}[\textit{:}\mbox{}]
	\titlespacing{\paragraph}{2pt}{-\sspace}{-\ssspace}
\fi

\graphicspath{ {./figures/} }

\begin{document}
	
\newgeometry{top=\mytopdis,left=\mydis,right=\mydis,bottom=\mydisbottom}
\title{Stochastic Resource Allocation for \\Electricity Distribution Network Resilience}	
	
	\author{Derek Chang, Devendra Shelar, and Saurabh Amin \\ Massachusetts Institute of Technology
		\thanks{The first two authors contributed equally to this work.}
		\thanks{Mailing address: Massachusetts Institute of Technology, 77 Massachusetts Avenue 1-241, Cambridge, MA 02139 USA (e-mail: \{changd,shelard,amins\}@mit.edu, phone: 857-253-8964).}
		\thanks{This work was supported by NSF CAREER award CNS 1453126, NSF Graduate Research Fellowship under Grant No. 1122374, and NSF FORCES award CNS-1239054.}
	}


	\maketitle
	
	\thispagestyle{empty}
	\begin{abstract}
		In recent years, it has become crucial to improve the resilience of electricity distribution networks (DNs) against storm-induced failures. Microgrids enabled by Distributed Energy Resources (DERs) can significantly help speed up re-energization of loads, particularly in the complete absence of bulk power supply. We describe an integrated approach which considers a pre-storm DER allocation problem under the uncertainty of failure scenarios as well as a post-storm dispatch problem in microgrids during the multi-period repair of the failed components. This problem is computationally challenging because the number of scenarios (resp. binary variables) increases exponentially (resp. quadratically) in the network size. Our overall solution approach for solving the resulting two-stage mixed-integer linear program (MILP) involves implementing the sample average approximation (SAA) method and Benders Decomposition. Additionally, we implement a greedy approach to reduce the computational time requirements of the post-storm repair scheduling and dispatch problem. The optimality of the resulting solution is evaluated on a modified IEEE 36-node network.
	\end{abstract}
	
	
	%
\section{Introduction} \label{sec:intro}
	In the U.S., weather-induced disruptions to power systems cost \$20-\$55 billion in annual economic losses~\cite{ExecReport}. Among these disruptions, about 90\% of outages occur in electricity distribution networks (DNs)~\cite{Campbell}. Smart grid technologies such as microgrids powered by Distributed Energy Resources (DERs) permit DNs to provide power to loads even when the bulk supply from central generation is disrupted~\cite{hurricaneMicrogrid, Badolato}. Current disaster preparedness procedures include use of microgrids for operational benefits~\cite{ShelarAminHiskens}. For example, before Hurricane Sandy struck, the Federal Emergency Management Agency prepared an inventory of industrial-size emergency generators~\cite{mobileEmergency}. However, to realize the benefit of DER-enabled microgrids, agencies need to ensure coordinated resource allocation and response actions. Otherwise, the available resources for supporting power dispatch and repair are likely to be ineffective in reducing prolonged outages and economic losses~\cite{Harvey, Irma}.
	
	
	DER allocation in microgrids has received attention in the context of remote control applied to microgrids~\cite{Chen}; allocation of mobile generators~\cite{mobileEmergency}; utilization of electric buses as generation resources~\cite{prehurricaneAllocation}; and allocation for microgrid formation in radial and meshed topologies~\cite{Lamadrid}. However, previous approaches do not also consider damage uncertainty and dynamic repair of damaged network components; this limitation can result in suboptimal resource utilization. 

		
	
\begin{figure}[htbp!]
		\centering
		\def \hhh {\textwidth}
		\ifOneColumn 
		\def \hhh {0.7\textwidth}
		\fi
		\resizebox{\hhh}{!}{
			\begin{tikzpicture}[scale=1.2]
			\draw [thick,-] (-1,0.15) -- (-1,-0.15);
			\node[align=center] at (-2.0,0.5) {Failure \\prediction \\($\probFails$)};
			
			\node[align=center] at (-2.0,-0.75) {Storm\\forecast};
			
			\node[align=center] at (0.25,0.5) {Allocation ($\first$)};
			
			\def \stormlanding {1.5}
			\def \hh {1.3}
			\def \yy {-0.6}
			\def \yyy {-0.95}
			\node[align=center] at (\stormlanding,\hh) {Storm\\landing};
			\draw [thick,-] (\stormlanding,0.1) -- (\stormlanding,-0.1);	
			\draw [thick,->] (\stormlanding,0.9) -- (\stormlanding,0.15);
			\node[align=center] at (2.9,0.5) {Line failures $(\scenarioIdx)$};
			\draw [thick,->] (4.25,0.9) -- (4.25,0.15);
			\node[align=center] at (4.25,1.3) {Storm\\passing};
			\draw [thick,-] (4.25,0.1) -- (4.25,-0.1);	
			\draw [thick,-] (4.5,0.1) -- (4.5,-0.1);	
			\draw [thick,-] (4.75,0.1) -- (4.75,-0.1);	
			\draw [thick,-] (5,0.1) -- (5,-0.1);	
			\draw [thick,-] (5.25,0.1) -- (5.25,-0.1);	
			\draw [thick,-] (5.5,0.1) -- (5.5,-0.1);	
			\draw [thick,-] (5.75,0.1) -- (5.75,-0.1);	
			\draw [thick,-] (6,0.1) -- (6,-0.1);	
			\draw [thick,-] (6.25,0.1) -- (6.25,-0.1);	
			\draw [thick,-] (6.5,0.1) -- (6.5,-0.1);
			\draw [thick,-] (6.75,0.1) -- (6.75,-0.1);
			\draw [thick,-] (7,0.1) -- (7,-0.1);	
			\draw [thick,-] (7.25,0.1) -- (7.25,-0.1);	
			\draw [thick,-] (7.5,0.1) -- (7.5,-0.1);	
			\draw [thick,-] (7.75,0.1) -- (7.75,-0.1);	
			\draw [thick,-] (8,0.1) -- (8,-0.1);	
			\draw [thick,-] (8.25,0.1) -- (8.25,-0.1);
			\node[align=center] at (6.25,0.6) {Line repairs ($\yc{}{}$)};
			\node[align=center] at (6.25,-0.6) {DER Dispatch ($\xc{}{}$) };
			\draw [thick,->] (4.25,\yy) -- (4.25,-0.15);
			\node[align=center] (k0) at (4.3, \yyy) {$\period=0$};
			
			\draw [thick,->] (8.25,\yy) -- (8.25,-0.15);
			\node[align=center] at (8.25,\yyy) {\\$\period=\nperiod$};
			\node[align=center] at (8.25,1.3) {Reconnect\\to grid};
			\draw [thick,->] (8.25,0.9) -- (8.25,0.15);
			\draw [thick,->] (-2.75,0) -- (8.5,0);
			\draw [thick,decorate,decoration={brace,amplitude=6pt,raise=0pt}] (4.5,0.15) -- (8.0,0.15);
			\draw [thick,decorate,decoration={brace,amplitude=6pt,raise=0pt}] (8.0,-0.15) -- (4.5,-0.15);
			\end{tikzpicture}}
		\caption{Timeline of events and decisions. $\protect\probFails$ denotes the distribution over failure scenarios; $\protect\first$ the pre-storm DER allocation decision; $\protect\scenarioIdx$ a realization of uncertainty; $\protect y$ the line repair schedule; $\protect x$ the network state variables. Line repair and dispatch decisions are undertaken over periods $\protect\period=0,1,\cdots,\nperiod$.} 
		\label{fig:timeline}
	\end{figure}
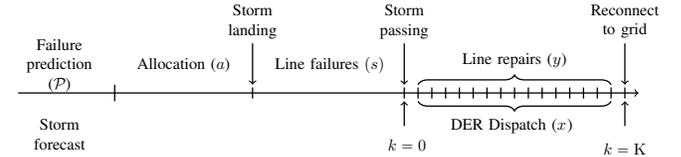

	\Cref{sec:twoStageProb} summarizes our modeling approach for improving resilience of DNs against tropical storms~\cite{SGC}. The approach jointly considers proactive pre-storm decisions (DER allocation) and post-storm response actions of microgrid formation, component repairs, and DER dispatch (see~\Cref{fig:timeline}). In contrast to~\cite{Chen,mobileEmergency,Lamadrid}, our model considers DN component repairs over a multi-period horizon. The model is written as a two-stage stochastic mixed-integer program (SMIP2).  
	
Our first contribution of this work is a stylized example that demonstrates how uncertainty in line failures, repair scheduling, and power flows can affect the optimal DER allocation (see \Cref{sec:example}). The presented example highlights the necessity of the added modeling complexity associated with SMIP2.

Our second contribution (\Cref{sec:solutionApproach}) is a solution approach based on the sample average approximation (SAA) method~\cite{Shapiro}, which involves restricting SMIP2 to a 
subset of scenarios, and solving the resulting MIP using L-shaped Benders Decomposition (LBD)~\cite{LshapedBendersDecomposition}. The approach significantly the decreases computation time to solve SMIP2, which is a computationally challenging problem because the number of scenarios (resp. the Stage II binary variables) increases exponentially (resp. quadratically) with the network size. 
 
 Although applying SAA and LBD results in a smaller MIP, LBD requires solving computationally expensive Stage II subproblems (typically solved to optimality using branch-and-bound algorithms). Our third contribution is a greedy heuristic that sequentially determines optimal repair actions in a period-wise manner. We show that this heuristic provides reasonable upper bounds to the Stage II subproblems, thus significantly reducing the computation time required to solve SMIP2.
 
 
 We evaluate our solution approach in \Cref{sec:compStudy}, and conclude our work in \Cref{sec:concludingRemarks}. 
 
    \restoregeometry\noindent

		\section{Two-stage stochastic program}\label{sec:twoStageProb}
		
			\begin{table}[htbp!]
				\centering
			
			\def \xx {0.5cm}
			\def \xxx {5.0cm}
			\def \xxxx {7.0cm}
			\def \xxxxx {1cm}
			\def \xxxxxx {10cm}
			\def \xxxxxxx {6.5cm}
			\ifOneColumn
			\def \ww {0.5\textwidth}
			\else
			\def \ww {\textwidth}
			\fi 
			\resizebox{\ww}{!}{
				\begin{tabular}{p{1.67cm}p{8cm}}
					\multicolumn{2}{l}{\textbf{DN  parameters}} \\
					\small		$\N$ & set of nodes in DN \\
					$\E$ & set of edges in DN \\
					$0$ & substation node label\\
					$\G$ & radial topology of DN, $\G = (\N\bigcup\{0\},\E)$ \\
					$\NN = \abs{\N}$ & number of non-substation nodes in DN\\
					$\setSite\subseteq \N$ & a set of potential locations for developing DER sites\\
					$\setDER$ & set of available DERs\\
					$\nuc{nom}{}$ & nominal squared voltage magnitude (1 pu)\\			
					\multicolumn{2}{l}{\textbf{Parameters of edge $e \in \E $}}  \\
					$\Pc{\period,\scenarioIdx}{e}, \Qc{\period,\scenarioIdx}{e} $ & active and reactive power flowing on line $e$ \\
					$\resistance{e}, \reactance{e}$ & resistance and reactance of line $(i,j) \in \E$  \\
					$\fromNode{e}, \toNode{e}$ & from and to nodes of line $e$ between nodes $i$ and $j$\\
					\multicolumn{2}{l}{\textbf{Nodal quantities of node $i \in \N $}} \\
					$\nuc{\period,\scenarioIdx}{i}$ & squared voltage magnitude at node $i$\\
					$\ptc{\period,\scenarioIdx}{i}, \qtc{\period,\scenarioIdx}{i}$ & net active and reactive power consumed at node $i$ 
					\\
					\multicolumn{2}{l}{\textbf{Quantities of load at node $i \in \N $}} \\
					$\pcc{max}{i}, \qcc{max}{i}$ & nominal active and reactive power demand at node $i$  \\
					$\kcc{\period,\scenarioIdx}{i}$ & 0 if load at node  $i$ is connected to DN; 1 otherwise \\		
					$\lcc{\period,\scenarioIdx}{i}$ & fraction of demand satisfied at node $i$\\
					$\lcc{min}{i}$ & lower bound of  load control parameter $\lcc{}{i}$\\
					$\Cload_i$ & cost of unit load control at node $i$\\
					$\Cshed_i$ & cost of load shedding at node $i$\\
					$\pcc{\period,\scenarioIdx}{i} , \qcc{\period,\scenarioIdx}{i}$ & actual active and reactive power consumed at node $i$  \\
					$\nucc{min}{i},\nucc{max}{i}$ & lower, upper voltage bounds for load  at node $i$\\
					$\Csite_i$ & cost of developing DER site at node $i$
					\\
					\multicolumn{2}{l}{\textbf{Quantities of DER $\der \in \setDER$}} \\
					$\pgc{max}{\der}, \qgc{max}{\der}$ & maximum active and reactive power bounds  of DER $\der$ \\
					$\pfc{max}{\der}$ & $\tan\arccos$ of the maximum power factor of DER $\der$\\
					$\pgc{\period,\scenarioIdx}{i\der}, \qgc{\period,\scenarioIdx}{i\der}$ & active and reactive power contribution of DER $\der$ at node $i$ \\
					$\kgc{\period,\scenarioIdx}{i}$ & 0 if DG at node $i$ is connected to DN; 1 otherwise \\
					$\nugc{min}{i},\nugc{max}{i}$ & lower, upper voltage bounds for DG  at node $i$\\
					$\kqc{\der}$ & voltage droop coefficient of the DER $\der$\\ $\nuc{ref}{\der}$ & idle (no load) voltage reference setpoint of DER~\cite{droopControl}\\ 
					\multicolumn{2}{l}{\textbf{Failure variables for scenario $\scenarioIdx$}} \\
					$\scenarioIdx \in \{0,1\}^{\E}$ & $\scenarioIdx_e =1$ if line $e$ is disrupted; 0 otherwise. \\
					$\E[\scenarioIdx]$ & set of lines failed in scenario $\scenarioIdx$. \\		
					\multicolumn{2}{l}{\textbf{Allocation decision variables}} \\
					$\ysc{}{}\in \setBinary^{\setSite}$ & $\ysc{}{i} = 1$ if a site is developed at node $i\in\setSite$; 0 otherwise. \\
					$\ygc{}{}\in \setBinary^{\setSite\times\setDER}$ & $\ygc{}{i\der} = 1$ if DER $\der$ is allocated at node $i\in\setSite$; 0 otherwise. \\
					\multicolumn{2}{l}{\textbf{Repair decision variables for scenario $\scenarioIdx$}} \\
					$\ylinec{\period,\scenarioIdx}{}\in\setBinary^{\E}$ & $\ylinec{\period,\scenarioIdx}{e} = 1$ if line $e$ is repaired in period $\period$; 0 otherwise. \\
					$\klinec{\period,\scenarioIdx}{}\in\setBinary^{\E}$ & $\klinec{\period,\scenarioIdx}{e} = 1$ if line $e$ is operational in period $\period$; 0 otherwise. \\ 		
			\end{tabular}}		
			\vspace{-3mm}	
			\label{tab:notations}
			\caption{Table of notation.} 
		\end{table}
	
	\vspace{1mm}
	Consider a DN denoted as $\G=(\N\cup\{0\},\E)$, where $0$ is the substation node, $\N$ the set of nodes, and $\E$ the set of edges. We formulate the SMIP2 problem as:
	\begin{align}\label{eq:sooFirstStage}\tag{SMIP2}
	{
		\begin{aligned}
	& \textstyle\min_{\first} \ && \soo{}{}(\first) \coloneqq \ \costFirstStage(\first) + \mathbb{E}_{ \scenario \sim \probFails} [\costSecondStage\left( \first, \scenario\right)], \\
	& \text{s.t.} && \firstStageConstraints\first\ge \firstStageRH, \quad \first \in \setBinary^{\na},
	\end{aligned}}
	\end{align}	
	where 
	$\first\in\setBinary^{\na}$ denotes a $\na$-dimensional binary vector that captures a DER allocation strategy; $\costFirstStage(\first)\coloneqq \costFirstStageCoeff^\top\first$ the Stage I cost; and $\firstStageConstraints\in \R^{\ma\times \na}, \firstStageRH\in\R^{\ma}$ model the set of feasible allocation strategies, where $\ma$ denotes the number of constraints on $\first$. 
	The distribution $\probFails$ characterizes the probability of line failures and is supported over $\setScenarios \coloneqq \{0,1\}^{\E}$; random vector $\scenario$ is drawn from $\probFails$. Finally, $\mathbb{E}_{\scenario\sim\probFails}[\costSecondStage(\first,\scenario)]$ denotes the expected Stage II cost under allocation $\first$.
	
	For a pair $(\first,\scenarioIdx)\in\First\times\setScenarios$,\  $\costSecondStage(\first,\scenarioIdx)$ denotes the optimal value of the Stage II recourse problem. Consider a multi-period horizon $\setPeriods = \{0,1,\cdots,\nperiod\}$ where each period is viewed as a work shift during which lines are repaired and DERs are dispatched. Then, we formulate the Stage II problem as a multi-period mixed-integer linear program (MILP) as follows:
	\begin{align}\label{eq:sooSecondStage}\tag{SP2}
		\begin{aligned}
	\hspace{-0.4cm}\costSecondStage\left( \first, \scenario\right) \coloneqq \mmin \quad & \ssum_{\period\in\setPeriods}  \costSecondStagePeriod{k}(\xc{\ts}{},\yc{\ts}{}) \\ \text{s.t.} \quad& \secondStageConstraints{}{x}\xc{\scenario}{} + \secondStageConstraints{}{y}\yc{\scenario}{} \ge \secondStageRh - \secondStageRhA\first \\
	& \xc{\ts}{} \in \R^{\nx},  \ \yc{\ts}{} \in \setBinary^{\nys} \ \forall\ \period\in\setPeriods  \\
	& \xc{\ts}{i} \in \{0,1\} \ \forall\ i\in\indexSet{x}, \period\in\setPeriods,
	\end{aligned}
	\end{align}	
	where $\xc{\ts}{}\in \setBinary^{\indexSet{x}}\times  \R^{[\nx]\backslash\indexSet{x}}$ and $\yc{\ts}{}\in \setBinary^{\nys}$ denote the mixed-binary network state variables and binary line repair actions in period $\period$ and scenario $\scenario$; $\indexSet{x} \subset \{1,2,\cdots,\nx\}$ the index set indicating  $\xc{\ts}{i}$ variables with binary restrictions; 
   $\yc{\scenarioIdx}{} = \{\yc{\ts}{}\}_{\period\in\setPeriods}$ the overall line repair schedule; $\xc{\scenarioIdx}{} = \{\xc{\ts}{}\}_{\period\in\setPeriods}$ the aggregated network state variables;  $\costSecondStagePeriod{k}(\xc{\ts}{},\yc{\ts}{})\coloneqq  \costSecondStageLin{\top}{\text{x},\period}\xc{\ts}{} + \costSecondStageLin{\top}{\text{y},\period}\yc{\ts}{}$ the Stage II cost; and $ \secondStageConstraints{}{\text{x}}\xc{\scenario}{} + \secondStageConstraints{}{\text{y}}\yc{\scenario}{} \ge \secondStageRh - \secondStageRhA\first$ the system of mixed-integer linear constraints on the Stage II decision variables. 

 In this paper, we assume the probability distribution  $\probFails$ as given. We refer the reader to~\cite{SGC} for details on estimating $\probFails$. 	We now describe our models for allocation ($\first$), repair ($\yc{\scenario}{}$), and dispatch  ($\xc{\scenario}{}$) actions, and the objective functions for both the stages; see Table I for a summary of notation. 
	
\subsection{Resource allocation model}\label{sec:derPlacement}
	Let $\setDER$ denote the set of available DERs, and $\setSite\subseteq\N$ the subset of nodes in which DERs can be feasibly allocated.
	In Stage I (before the storm), the operator needs to decide which sites to develop, and which DERs to allocate to the chosen sites.\footnote{DER site development, such as land acquisition, building enclosures and elevated platforms, ensure secure and reliable operation of DERs.} The Stage I constraints are as follows:
	\begin{subequations}
			\begin{alignat}{8}
			\label{eq:siteConditionNecessary}
			\ysc{}{i} &\le \ssum_{\der\in\setDER} \ygc{}{i\der} \qquad &&\forall \ \site \in\setSite\\
			\label{eq:siteConditionSufficient}
			\ygc{}{i\der} & \le \ysc{}{i}\qquad &&\forall \ \der\in\setDER, \site \in\setSite\\
			\label{eq:oneNodeCondition}
			\ssum_{i\in\setSite} \ygc{}{i\der} &\le 1\qquad &&\forall \ \der\in\setDER\\
			\label{eq:genResourceConstraint}
			\ssum_{i\in\setSite} \ssum_{\der\in\setDER} \ygc{}{i\der} &\le \genBudget,
		\end{alignat}
	\end{subequations}	
\noindent	where \eqref{eq:siteConditionNecessary} denotes that a site $\site$ is operational if there is at least one DER allocated to that site; a DER can be allocated to a site only if that site is developed~\eqref{eq:siteConditionSufficient}; a DER $\der$ can be allocated to at most one site \eqref{eq:oneNodeCondition}; and the total number of allocated DERs can be at most $\genBudget$  \eqref{eq:genResourceConstraint}. Here $\genBudget \le \abs{\setDER}$ models the supply constraint on the number of DERs.
	
	Thus, the Stage I decision variable (joint site development and DER allocation) in \eqref{eq:sooFirstStage} is defined as $\first\coloneqq \left(\ysc{}{},\ygc{}{}\right)$. The set of feasible resource allocation strategies is defined as $\First \coloneqq \{\left(\ysc{}{},\ygc{}{}\right) \in \setBinary^{\setSite}\times\setBinary^{\setSite\times\setDER} \ | \ \eqref{eq:siteConditionNecessary}-\eqref{eq:genResourceConstraint} \text{ hold} \}$. 

		\subsection{Multi-period joint repair scheduling and dispatch  model}\label{sec:repairModel}
		From a practical viewpoint, each period in the second-stage multi-period horizon can be viewed as one work shift of the repair crews. We assume that at period $\period = 0$, the DN is disconnected from the main grid due to the storm. Subnetworks formed as a result of line failures can be operated as microgrids using the available DER supply. As the line repairs continue over subsequent periods, smaller microgrids increase in size and/or merge together to form larger microgrids. At period $\period = \nperiod$, all line repairs are complete, the DN is reconnected to the main grid, and normal operation is restored. 
		
		We assume that the estimated number of periods to repair all failed lines $\nperiod=\abs{\E}$, for a straightforward comparison between different scenarios and repair crew constraints. This is not a restrictive assumption because if the repairs finish at period $\period < \nperiod - 1$, the network state will remain unchanged until $\period = \nperiod$, when normal operation is restored. 
		
		The constraints governing line repair decisions are: 
	{
		\begin{subequations}
			\begin{alignat}{8}
				\label{eq:norepairFirstPeriod}
				\ylinec{\period,\scenarioIdx}{e} &= 0 \quad &&\forall\  e \in \E[s], \period = 0\\
				\label{eq:lineRepairCrewConstraint}
				\ssum_{e\in\E[s]}\ \ylinec{\period,\scenarioIdx}{e} &\le \crewBudget \qquad &&\forall\ \period\in\setPeriods\setminus\nperiod\\		
				\label{eq:mainGridConnected}
				\ylinec{\period,\scenarioIdx}{e} &= 1 \quad &&\forall \ e \in\E[s], \fromNode{e} = 0, \period=\nperiod \\	
				\label{eq:norepair}
				\ssum_{\period=0}^\nperiod \ylinec{\period,\scenarioIdx}{e} &\leq 1 \quad &&\forall\ e\in\E[s]\\											
				\label{eq:initialLineDamage}
				\klinec{\period,\scenarioIdx}{e} &= 1 \qquad&&\forall\ e \in\E[s],  \period  = 0\\					
				\label{eq:lineDisconnect}
				\klinec{\period,\scenarioIdx}{e} &= 1 \quad &&\forall\ e, \period\in\setPeriods\setminus\nperiod, \fromNode{e}=0\\																		
				\label{eq:undamagedLines}
				\klinec{\ts}{e} &= 0\quad &&\forall\ \period\in \setPeriods, e\notin\E[s]\\
				\label{eq:lineRepairChange}
				\klinec{\period-1,\scenarioIdx}{e} - \ylinec{\period,\scenarioIdx}{e}	 &= \klinec{\period,\scenarioIdx}{e} \quad&&\forall\ \period \in \setPeriods\setminus 0, e\in\E[s] \\
			\label{eq:repairBinaryConstraint}
			\ylinec{\period,\scenarioIdx}{e}, \klinec{\ts}{e} &\in \setBinary \quad &&\forall\  e \in \E. 				
			\end{alignat}
		\end{subequations}
	}%
	No repairs are permitted at $\period = 0$ \eqref{eq:norepairFirstPeriod}; at most $\crewBudget$ lines can be repaired per period in $\period\in\{1,...,\nperiod-1\}$~\eqref{eq:lineRepairCrewConstraint};\footnote{The number of repairs can vary across periods depending on the number of crews. For simplicity, we assume that $\crewBudget$ is fixed at all periods.  
	} and the DN is reconnected to the main grid at $\nperiod$ \eqref{eq:mainGridConnected}, where $\fromNode{e} = 0$ denotes that line $e$ connects the DN to the substation node $0$.\footnote{Our model can be extended to allow early reconnection back to the main grid even before DN repairs are completed as shown in~\cite{shelarAminHiskensPart2}.} \Cref{eq:norepair} ensures a line can be repaired at most once. \Cref{eq:initialLineDamage} enforces that the damaged lines are non-operational at $\period = 0$; \eqref{eq:lineDisconnect} models that the line connected to the substation is non-operational until $\period = \nperiod$; and \eqref{eq:undamagedLines} captures that lines not damaged remain operational at all periods. A failed line turns operational after it is repaired~\eqref{eq:lineRepairChange}. The line repair actions and operating state variables are constrained to be binary 	\eqref{eq:repairBinaryConstraint}.
	
	The line repair variable for each scenario $\scenarioIdx$ is denoted as $\yc{\scenarioIdx}{} \coloneqq \{(\ylinec{\ts}{e}, \klinec{\ts}{e})\}_{e\in\E,\period\in\setPeriods}$, and the set of feasible repair schedules is  $\Yc{}{}\left(\scenarioIdx\right) \coloneqq \{\yc{}{} \in \E\times\setPeriods \ | \ \eqref{eq:norepairFirstPeriod}-\eqref{eq:repairBinaryConstraint} \text{ holds} \}$.\footnote{Our model can be easily extended to consider meshed topologies and network reconfiguration capabilities as in~\cite{Chen,Lamadrid}.}

Henceforth, we drop the notation $\forall\ \period\in\setPeriods, \scenarioIdx \in\setScenariosSmall$.

	In each period, the DERs are redispatched to satisfy new operating constraints resulting from lines becoming operational and to enable further load restoration. The following constraints characterize our DER model: 
{
	\begin{subequations}
		\begin{alignat}{8}
			\label{eq:derSiteActiveContribution} 0 \le \pgc{\ts}{i\der} &\le \ygc{}{i\der} \pgc{max}{\der} && \forall\ \der \in \setDER, i \in \setSite\\
			\label{eq:derSiteReactiveContribution} \abs{\qgc{\ts}{i\der}} &\le \pfc{max}{\der} \pgc{\ts}{i\der} && \forall\ \der \in \setDER,  i \in \setSite\\
			\label{eq:derNonSiteContribution}\pgc{\ts}{i\der} &= \qgc{\ts}{i\der} = 0 \qquad &&\forall\  \der \in \setDER, i\in\N\backslash\setSite\\
		\hspace{-0.3cm}	|\nuc{\ts}{i} - (\nuc{ref}{\der} &- \kqc{\der}\qgc{\ts}{i\der})| &&\le (1-\ygc{}{i\der})\bigM \nonumber\\
			\quad 
			\label{eq:islandVoltDroop}&&&\forall\  \der \in \setDER, i\in\setSite,  \period\in\setPeriods\backslash\nperiod, 
		\end{alignat}
	\end{subequations}
}%
where \eqref{eq:derSiteActiveContribution} bounds the active power contributed by a DER; \eqref{eq:derSiteReactiveContribution} models a power factor constraint, \eqref{eq:derNonSiteContribution} ensures no active and reactive power contributions of a DER to non-DER site nodes; and \eqref{eq:islandVoltDroop} models voltage droop control.\footnote{Once the DN is connected to the bulk grid, the \enquote{stiff} AC system of the bulk grid determines the terminal voltage of the DERs. Hence the voltage droop equation does not apply at period $\nperiod$.}

 The constraints governing our load model are as follows:\vspace{-0.4cm}
{
	\begin{subequations}
	\begin{alignat}{8}
		\label{eq:voltageDisconnectLoads}	\kcc{\ts}{i}  \ge \nucc{min}{i} - \nuc{\ts}{i}, &
	\ && \kcc{\ts}{i} \ge \nuc{\ts}{i} - \nucc{max}{i} \hspace{0.3cm}&& \forall\ i\in \N \\
		\label{eq:loadControlParameterConstraint}
		\lcc{\ts}{i} \ge (1-\kcc{\ts}{i})\lcc{min}{i}, & &&\lcc{\ts}{i}\le  (1-\kcc{\ts}{i})  &&\forall \ i\in\N\\
		\label{eq:loadControlEquation} 
		\pcc{\ts}{i} =  \lcc{\ts}{i}\ \pcc{max}{i}, & \ 
		&& \qcc{\ts}{i} =  \lcc{\ts}{i}\ \qcc{max}{i}  &&\forall\ i \in \N \\
	\label{eq:loadConnectivityBinaryConstraint}	& && \kcc{\ts}{i}\in \setBinary &&  \forall\ i\in \N. 	 
	\end{alignat}
	\end{subequations}}
Here, \eqref{eq:voltageDisconnectLoads} ensures that the load remains connected only if voltage bounds are satisfied; \eqref{eq:loadControlParameterConstraint} models bounds on load control; \eqref{eq:loadControlEquation} determines the load's active and reactive power consumption; and \eqref{eq:loadConnectivityBinaryConstraint} models a binary constraint on $\kcc{\ts}{i}$.

\vspace{0.05cm}
For computational simplicity, the power flow model is\\ \vspace{0.2cm}
	given by the \textit{LinDistFlow} model adapted to microgrids~\cite{shelarAminHiskensPart2}:
\vspace{-0.4cm}
		\begin{subequations}		
			\begin{alignat}{8}
				\label{eq:totalActivePowerConsumption} \ptc{\ts}{i} &= \pcc{\ts}{i} -  \ssum_{\der\in\setDER}\pgc{\ts}{i\der}  \qquad &&\forall\ i\in\N\\
				\label{eq:totalReactivePowerConsumption}\qtc{\ts}{i} &= \qcc{\ts}{i} -  \ssum_{\der\in\setDER}\qgc{\ts}{i\der} \qquad  &&\forall\ i\in\N\\
				\label{eq:activePowerConservation} \Pc{\ts}{e} &= \displaystyle\ssum_{l: \fromNode{l} = \toNode{e}} \Pc{\ts}{l} + \ptc{\ts}{\toNode{e}} \quad&& \forall\ e \in \E \\
				\label{eq:reactivePowerConservation}  \Qc{\ts}{e} &= \displaystyle\ssum_{l: \fromNode{l} = \toNode{e}}\Qc{\ts}{l} + \qtc{\ts}{\toNode{e}} && \forall\ e \in \E\\
				\label{eq:activePowerIsland}
				\abs{\Pc{\ts}{e}} & \le (1-\klinec{\ts}{e})\bigM \qquad && \forall\ e\in\E\\
				\label{eq:reactivePowerIsland} \abs{\Qc{\ts}{e}} & \le (1-\klinec{\ts}{e})\bigM \qquad && \forall\ e\in\E
			\end{alignat}	
		\begin{equation}
						\label{eq:voltageDrop} \hspace{-0.3cm}	 {\small |\nuc{\period,\scenarioIdx}{\toNode{e}} -  (\nuc{\period,\scenarioIdx}{\fromNode{e}}-  2(\resistance{e}\Pc{\period,\scenarioIdx}{e} +  \reactance{e}\Qc{\period,\scenarioIdx}{e}))| \le \bigM\klinec{\period,\scenarioIdx}{e}\ \forall \ e \in \E.}
		\end{equation}
		\end{subequations}
	Eqs.~\eqref{eq:totalActivePowerConsumption}-\eqref{eq:totalReactivePowerConsumption} determine the net active and reactive power consumed at the nodes; \eqref{eq:activePowerConservation}-\eqref{eq:reactivePowerConservation} compute the resulting active and reactive power flows on the lines; \eqref{eq:activePowerIsland}-\eqref{eq:reactivePowerIsland} ensure that no power flows on the failed lines until they are repaired; and \eqref{eq:voltageDrop} ensures that the voltage drop constraint along a line $e$ (between `from' node $\fromNode{e}$ and `to' node $\toNode{e}$) is enforced only if $e$ is operational.\footnote{When the DN is connected back to the main grid, the substation voltage is assumed to be the nominal voltage.}
	
	The dispatch variable for scenario $s$ is denoted as $\xc{\scenarioIdx}{} \coloneqq$  $\{\pgc{\ts}{},\qgc{\ts}{},$ $\lcc{\ts}{},\kcc{\ts}{},\ptc{\ts}{},\qtc{\ts}{},\Pc{\ts}{},\Qc{\ts}{},$ $\nuc{\ts}{}\}_{\period\in\setPeriods}$.
	The set of feasible power flows under allocation $\first$ and line repair schedule $\yc{\scenarioIdx}{}\in\Yc{}{}\left(\scenarioIdx\right)$ is written as $\Xc{}{}\left(\first,\scenarioIdx,\yc{\scenario}{}\right) \coloneqq \{\xc{}{}\ |\  \eqref{eq:derSiteActiveContribution}-\eqref{eq:voltageDrop} \text{ hold}\}$. $\Xc{}{}\left(\first,\scenarioIdx,\yc{\scenarioIdx}{}\right)$. The sets $\Xc{}{}\left(\first,\scenarioIdx,\yc{\scenario}{}\right)$ and $\Yc{}{}\left(\scenarioIdx\right)$ define the system of inequalities $\secondStageConstraints{}{x}\xc{\scenario}{} + \secondStageConstraints{}{y}\yc{\scenario}{}  = \secondStageRh - \secondStageRhA\first$ and the binary constraints in \eqref{eq:sooSecondStage}.

\subsection{Objectives} \label{subsec:objective}
We assume that the Stage I cost is dominated by the site development cost. Thus, the Stage I cost is given as $\costFirstStage(a) = \sum_{i\in\setSite}\Csite_i \ysc{}{i}$, i.e., the DER allocation has zero cost. For the Stage II objective, we assume costs are the same at all periods, i.e., $\costSecondStageLin{}{x,j} =  \costSecondStageLin{}{x,k} \ \forall\ j, k\in\setPeriods$. Furthermore, we assume that $\costSecondStageLin{}{y,\period} = \mathbf{0}\ \forall\ \period\in\setPeriods$, i.e.,  there is no cost of line repairs. (The model can be easily extended to account for objectives without these assumptions.) Let the cost of load control/shedding of a load at node $i\in\N$ be defined as:
\begin{equation}\label{eq:costLoad}
	\costLoad{i}(\kcc{\ts}{i},\lcc{\ts}{i}) = \Cload_i(1-\lcc{\ts}{i}) +  (\Cshed_i-\Cload_i)\kcc{\ts}{i}. 
\end{equation}
We can define the value or benefit to the operator by operating a load $i$ as \begin{equation}\label{eq:valueLoad}
	\valueSecondStagePeriod{i}(\kcc{\ts}{i},\lcc{\ts}{i}) \coloneqq \Cshed_i-\costLoad{i}(\kcc{\ts}{i},\lcc{\ts}{i}). 
\end{equation}
Then, we define the Stage II objective function to be the weighted sum of the cost of load control and load shedding, specifically:
	 	$\ \costSecondStagePeriod{\period}(\xc{\ts}{},\yc{\ts}{}) = \sum_{\node\in\N}  \costLoad{i}(\kcc{\ts}{i},\lcc{\ts}{i}).$

\vspace{0.1cm}
Thus, the DEF reformulation of \eqref{eq:sooFirstStage} is posed as:
\begin{align}
\begin{aligned} \label{eq:sooResourceAlloc}
\min_{\first,\xc{}{},\yc{}{}} \quad & 
	\sum_{i\in\setSite} \text{W}^{\text{SD}}_{i} \ysc{}{i} +
	\sum_{\scenario\in\setScenarios} \probFails(\scenario) \sum_{\period \in \setPeriods} \costSecondStagePeriod{\period}(\xc{\ts}{},\yc{\ts}{})\\
\text{s.t.  }\quad &\first\in\First, \yc{\scenarioIdx}{}\in\Yc{}{}\left(\scenarioIdx\right), \xc{\scenarioIdx}{} \in \Xc{}{}\left(\first,\scenarioIdx,\yc{\scenarioIdx}{}\right) \  \forall \ \scenarioIdx\in\setScenarios,
\end{aligned}
\end{align}
	where $\xc{}{} \coloneqq \{\xc{\scenarioIdx}{}\}_{\scenarioIdx\in\setScenarios}$ and $\yc{}{} \coloneqq \{\yc{\scenarioIdx}{}\}_{\scenarioIdx\in\setScenarios}$.

For a period $\period$, we also define the system performance metric $\systemPerformance$ as follows:
\begin{equation}
\label{eq:SP}
\hspace{-0.2cm}\systemPerformance = 
\frac{1}{\abs{\setScenarios}}\sum_{\scenarioIdx \in \setScenarios} 100\left(1-\costSecondStagePeriod{\period}({\xc{\ts}{}}^\star,{\yc{\ts}{}}^\star)/\big(\ssum_{i\in\N}\Cshed_i\big)\right)
\end{equation}	
where ${\xc{\scenarioIdx}{}}^\star\coloneqq\{{\xc{\ts}{}}^\star\}_{\period\in\setPeriods}$, ${\yc{\scenarioIdx}{}}^\star\coloneqq\{{\yc{\ts}{}}^\star\}_{\period\in\setPeriods}$ are the optimal solutions to  $\costSecondStage(\first, \scenarioIdx)$. System performance decreases with increasing costs  $\costSecondStagePeriod{\period}$, and is a maximum of 100 when the demand is fully met. 

\section{An illustrative example} \label{sec:example}

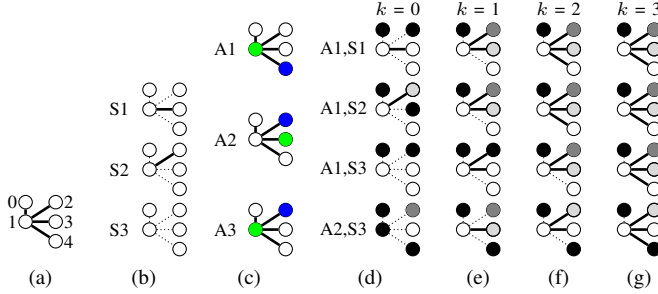
\begin{figure}[htbp!]
	\centering
	\tikzset{every node/.append ={font size=tiny}} 
	\def \ms {0.3}
	\def \is {1.8}
	\tikzstyle{dnnode}=[draw,circle, minimum size=\ms pt, inner sep = \is]
	\tikzstyle{dnedge}=[-, line width=1pt]
	\tikzstyle{vuledge}=[-, line width=3pt, red!70]
	\tikzstyle{swind}=[->, line width=2pt, blue!25, >=latex']
	\tikzstyle{dernode}=[circle, fill=blue, minimum size=\ms pt, inner sep = \is]
	\tikzstyle{dernodeLarge}=[circle, fill=green, minimum size=\ms pt, inner sep = \is]
	\tikzstyle{blackoutnode}=[circle, fill=black, minimum size=\ms pt, inner sep = \is]
	\tikzstyle{graynode}=[draw=black, pattern=north west lines, fill = gray, circle,  minimum size=\ms pt, inner sep = \is, fill opacity=0.3]
	\tikzstyle{graynodes}=[circle, pattern=north west lines, fill = gray, minimum size=\ms pt, inner sep = \is,fill opacity = 1]
	\tikzstyle{failededge}=[-, densely dotted]
	\def \drawgrid {\draw[step=1,gray, ultra thin, draw opacity = 0.5] (0,0) grid (3,4);}
	\def \drawSubstation {\draw[-, line width = 2pt] (0.8,4.05) -- (2.2,4.05)  node [midway,above] {};}
	
	\def \hx {0}
	\def \hy {0}
	\def \hyzero {0}
	\def \hytop {2}
	\def \hybelow {-2}
	\def \hylow {-4}
	
 	\def \drawZero {\node[dnnode] (0) at (\hx+1.5,\hy+0.65) {};}
	\def \drawOne {\node[dnnode] (1) at (\hx+1.5,\hy+0) {};}
	\def \drawTwo {\node[dnnode] (2) at (\hx+2.5,\hy+0.65) {};}
	\def \drawThree {\node[dnnode] (3) at (\hx+2.5,\hy+0) {};}
	\def \drawFour {\node[dnnode] (4) at (\hx+2.5,\hy-0.65) {};}
	
	\def \myfont {\scriptsize}
	\def \drawScenario (#1) {\node[] (0) at (\hx+0.5,\hy+0) {\myfont S#1};}
	\def \drawAllocation (#1) {\node[] (0) at (\hx+0.5,\hy+0) {\myfont A#1};}
	\def \drawDescription (#1,#2) {\node[] (0) at (\hx+0.2,\hy+0) {\myfont A#1,S#2};}
	\def \drawTime (#1) {\node[] (time) at (\hx+2,\hy+3.35) {\myfont $\period$ = #1};}
	\def \myscale {0.4}
	
	\subfloat[]{\label{subfig:nominal}
		\begin{tikzpicture}[scale=\myscale,every node/.append ={font size=tiny}]
		
		\def \hx {0}
		\def \hy {2}
		\drawZero \drawOne \drawTwo \drawThree \drawFour
		
		\foreach \from/\to in {0/1, 1/2, 1/3, 1/4}
		\draw[dnedge] (\from) -- (\to);		
		
		\node[xshift=-5] at (0) {\myfont 0};
		\node[xshift=-5] at (1) {\myfont 1}; 
		\node[xshift=5] at (2) {\myfont 2}; 
		\node[xshift=5] at (3) {\myfont 3};
		\node[xshift=5] at (4) {\myfont 4};
		\end{tikzpicture}
	}
	\subfloat[]{\label{subfig:scenarios}
		\begin{tikzpicture}[scale=\myscale,every node/.append ={font size=tiny}]
		
		\def \hx {0}
		\def \hy {2}
		\drawZero \drawOne \drawTwo \drawThree \drawFour
		\foreach \from/\to in {0/1, 1/2, 1/4}
		\draw[failededge] (\from) -- (\to);		
		\foreach \from/\to in {1/3}
		\draw[dnedge] (\from) -- (\to);	
		\drawScenario(1)
		
		\def \hx {0}
		\def \hy {0}
		\drawZero \drawOne \drawTwo \drawThree \drawFour
		\foreach \from/\to in {0/1, 1/3, 1/4}
		\draw[failededge] (\from) -- (\to);		
		\foreach \from/\to in {1/2}
		\draw[dnedge] (\from) -- (\to);	
		\drawScenario(2)
		
		\def \hx {0}
		\def \hy {-2}
		\drawZero \drawOne \drawTwo \drawThree \drawFour
		\foreach \from/\to in {0/1, 1/2, 1/3, 1/4}
		\draw[failededge] (\from) -- (\to);		
		\drawScenario(3)
		\end{tikzpicture}
	}
	\subfloat[]{\label{subfig:allocations}
	\begin{tikzpicture}[scale=\myscale,every node/.append ={font size=tiny}]
	
	\def \hx {0}
	\def \hy {\hyzero}
	\drawZero \drawOne \drawTwo \drawThree \drawFour
	
	\foreach \from/\to in {0/1, 1/2, 1/3, 1/4}
	\draw[dnedge] (\from) -- (\to);		
	
	\node[dernode] at (4) {} ;
	\node[dernodeLarge] at (1) {} ;
	\drawAllocation(1)
	
		\def \hx {0}
	\def \hy {-3}
	\drawZero \drawOne \drawTwo \drawThree \drawFour
	
	\foreach \from/\to in {0/1, 1/2, 1/3, 1/4}
	\draw[dnedge] (\from) -- (\to);		
	
	\node[dernode] at (2) {} ;
	\node[dernodeLarge] at (3) {} ;
	\drawAllocation(2)
	
		\def \hx {0}
	\def \hy {-6}
	\drawZero \drawOne \drawTwo \drawThree \drawFour
	
	\foreach \from/\to in {0/1, 1/2, 1/3, 1/4}
	\draw[dnedge] (\from) -- (\to);		
	
	\node[dernode] at (2) {} ;
	\node[dernodeLarge] at (1) {} ;
	\drawAllocation(3)
	\end{tikzpicture}
}
	\subfloat[]{\label{subfig:restorationTime0}
		\begin{tikzpicture}[scale=\myscale]
			\def \hx {0}
		\def \hy {\hytop}
		\drawZero \drawOne \drawTwo \drawThree \drawFour
		\foreach \from/\to in {0/1, 1/2, 1/4}
		\draw[failededge] (\from) -- (\to);		
		\foreach \from/\to in {1/3}
		\draw[dnedge] (\from) -- (\to);	
		\foreach \bn in {0,2}
		\node[blackoutnode] (\bn) at (\bn) {};
		\drawDescription(1,1)
		
		\def \hx {0}
		\def \hy {\hyzero}
		\drawTime(0)
		\drawZero \drawOne \drawTwo \drawThree \drawFour
		\foreach \from/\to in {0/1, 1/3, 1/4}
		\draw[failededge] (\from) -- (\to);		
		\foreach \from/\to in {1/2}
		\draw[dnedge] (\from) -- (\to);	
		\foreach \bn in {0,3}
		\node[blackoutnode] (\bn) at (\bn) {};
		\node[graynode] (2) at (2) {} ;
		\drawDescription(1,2)
		
		\def \hx {0}
		\def \hy {\hybelow}
		\drawZero \drawOne \drawTwo \drawThree \drawFour
		\foreach \from/\to in {0/1, 1/2, 1/3, 1/4}
		\draw[failededge] (\from) -- (\to);				
		\foreach \bn in { 0,2}
		\node[blackoutnode] (\bn) at (\bn) {};				
		\foreach \from/\to in {0/0}
		\draw[dnedge] (\from) -- (\to);		
		\drawDescription(1,3)
	
		\def \hx {0}
		\def \hy {\hylow}
		\drawZero \drawOne \drawTwo \drawThree \drawFour
		\foreach \from/\to in {0/1, 1/2, 1/3, 1/4}
		\draw[failededge] (\from) -- (\to);		
		
		\foreach \bn in {0, 1,4}
		\node[blackoutnode] (\bn) at (\bn) {};		
		\foreach \from/\to in {0/0}
		\draw[dnedge] (\from) -- (\to);		
		\node[graynodes] (2) at (2) {} ;
		\drawDescription(2,3)
		\end{tikzpicture}
	}
	\subfloat[]{\label{subfig:restorationTime1}
		\begin{tikzpicture}[scale=\myscale]
		
		\def \hx {0}
	\def \hy {\hytop}
	\drawZero \drawOne \drawTwo \drawThree \drawFour
	\foreach \from/\to in {0/1, 1/2, 1/4}
	\draw[failededge] (\from) -- (\to);		
	\foreach \from/\to in {1/3,1/2}
	\draw[dnedge] (\from) -- (\to);	
	\foreach \bn in {0}
	\node[blackoutnode] (\bn) at (\bn) {};
	\node[graynodes] (2) at (2) {} ;
	\node[graynode] (3) at (3) {} ;
	
	\def \hx {0}
	\def \hy {\hyzero}
	\drawTime(1)
	\drawZero \drawOne \drawTwo \drawThree \drawFour
	\foreach \from/\to in {0/1, 1/3, 1/4}
	\draw[failededge] (\from) -- (\to);		
	\foreach \from/\to in {1/2,1/3}
	\draw[dnedge] (\from) -- (\to);	
	\foreach \bn in {0}
	\node[blackoutnode] (\bn) at (\bn) {};
	\node[graynodes] (2) at (2) {} ;
	\node[graynode] (3) at (3) {} ;

	\def \hx {0}
	\def \hy {\hybelow}
	\drawZero \drawOne \drawTwo \drawThree \drawFour
	\foreach \from/\to in {0/1, 1/2, 1/3, 1/4}
	\draw[failededge] (\from) -- (\to);		
	\foreach \bn in { 0, 2}
	\node[blackoutnode] (\bn) at (\bn) {};
	\foreach \from/\to in { 1/3,1/2}
	\draw[dnedge] (\from) -- (\to);		
	\foreach \bn in {0,2}
	\node[blackoutnode] (\bn) at (\bn) {};

	\def \hx {0}
	\def \hy {\hylow}
	\drawZero \drawOne \drawTwo \drawThree \drawFour
	\foreach \from/\to in {0/1, 1/2, 1/3, 1/4}
	\draw[failededge] (\from) -- (\to);		
	\foreach \from/\to in { 3/1}
	\draw[dnedge] (\from) -- (\to);		
	\foreach \bn in {0,4}
	\node[blackoutnode] (\bn) at (\bn) {};	
		\node[graynodes] (2) at (2) {} ;
		\node[graynode] (3) at (3) {} ;
		\end{tikzpicture}
	}
	\subfloat[]{\label{subfig:restorationTime2}
	\begin{tikzpicture}[scale=\myscale]
	
	\def \hx {0}
	\def \hy {\hytop}
	\drawZero \drawOne \drawTwo \drawThree \drawFour
	\foreach \from/\to in {0/1, 1/2, 1/4}
	\draw[failededge] (\from) -- (\to);		
	\foreach \from/\to in {1/3,1/2,1/4}
	\draw[dnedge] (\from) -- (\to);	
	\foreach \bn in {0}
	\node[blackoutnode] (\bn) at (\bn) {};
	\node[graynodes] (2) at (2) {} ;
	\node[graynode] (3) at (3) {} ;
	
	\def \hx {0}
	\def \hy {\hyzero}
	\drawTime(2)
	\drawZero \drawOne \drawTwo \drawThree \drawFour
	\foreach \from/\to in {0/1, 1/3, 1/4}
	\draw[failededge] (\from) -- (\to);		
	\foreach \from/\to in {1/2,1/3,1/4}
	\draw[dnedge] (\from) -- (\to);	
	\foreach \bn in {0}
	\node[blackoutnode] (\bn) at (\bn) {};
	\node[graynodes] (2) at (2) {} ;
	\node[graynode] (3) at (3) {} ;

	\def \hx {0}
	\def \hy {\hybelow}
	\drawZero \drawOne \drawTwo \drawThree \drawFour
	\foreach \from/\to in {0/1, 1/2, 1/3, 1/4}
	\draw[failededge] (\from) -- (\to);		
	\foreach \bn in { 0}
	\node[blackoutnode] (\bn) at (\bn) {};
	\foreach \from/\to in { 1/3,1/2}
	\draw[dnedge] (\from) -- (\to);		
		\node[graynodes] (2) at (2) {} ;
		\node[graynode] (3) at (3) {} ;
	
	\def \hx {0}
	\def \hy {\hylow}
	\drawZero \drawOne \drawTwo \drawThree \drawFour
	\foreach \from/\to in {0/1, 1/2, 1/3, 1/4}
	\draw[failededge] (\from) -- (\to);		
	\foreach \from/\to in { 1/2,1/3}
	\draw[dnedge] (\from) -- (\to);		
	\foreach \bn in {0,4}
	\node[blackoutnode] (\bn) at (\bn) {};	
	\node[graynode] (2) at (2) {} ;
	\end{tikzpicture}
}
	\subfloat[]{\label{subfig:restorationTime3}
	\begin{tikzpicture}[scale=\myscale]
	
	\def \hx {0}
	\def \hy {\hytop}
	\drawZero \drawOne \drawTwo \drawThree \drawFour
	\foreach \from/\to in {0/1, 1/2, 1/4}
	\draw[failededge] (\from) -- (\to);		
	\foreach \from/\to in {1/3,1/2,1/4}
	\draw[dnedge] (\from) -- (\to);	
	\foreach \bn in {0}
	\node[blackoutnode] (\bn) at (\bn) {};
	\node[graynodes] (2) at (2) {} ;
	\node[graynode] (3) at (3) {} ;
	
	\def \hx {0}
	\def \hy {\hyzero}
	\drawTime(3)
	\drawZero \drawOne \drawTwo \drawThree \drawFour
	\foreach \from/\to in {0/1, 1/3, 1/4}
	\draw[failededge] (\from) -- (\to);		
	\foreach \from/\to in {1/2,1/3,1/4}
	\draw[dnedge] (\from) -- (\to);	
	\foreach \bn in {0}
	\node[blackoutnode] (\bn) at (\bn) {};
	\node[graynodes] (2) at (2) {} ;
	\node[graynode] (3) at (3) {} ;

	\def \hx {0}
	\def \hy {\hybelow}
	\drawZero \drawOne \drawTwo \drawThree \drawFour
	\foreach \from/\to in {0/1, 1/2, 1/3, 1/4}
	\draw[failededge] (\from) -- (\to);		
	\foreach \bn in { 0}
	\node[blackoutnode] (\bn) at (\bn) {};
	\foreach \from/\to in { 1/3,1/2,1/4}
	\draw[dnedge] (\from) -- (\to);		
	\node[graynodes] (2) at (2) {} ;
	\node[graynode] (3) at (3) {} ;
	
	\def \hx {0}
	\def \hy {\hylow}
	\drawZero \drawOne \drawTwo \drawThree \drawFour
	\foreach \from/\to in {0/1, 1/2, 1/3, 1/4}
	\draw[failededge] (\from) -- (\to);		
	\foreach \from/\to in { 1/2,1/3,1/4}
	\draw[dnedge] (\from) -- (\to);		
	\foreach \bn in {0,4}
	\node[blackoutnode] (\bn) at (\bn) {};	
	\node[graynode] (2) at (2) {} ;
	\end{tikzpicture}
}

	\caption{The subfigures show (a) nominal DN, (b) considered scenarios with failed lines shown by dotted lines (c) three potential DER allocations, (d) network topology after the storm, and (e)-(g) network restoration. Greater load control is indicated by a darker (grayer) node.}
	\label{fig:Schematic}
\end{figure}

In this section, we introduce an illustrative example to discuss how failure uncertainty, repair scheduling and power flow constraints affect the DER allocation (see \Cref{fig:Schematic}). 

\begin{table}[htbp!]\setlength\tabcolsep{2pt}
	\resizebox*{\textwidth}{!}{\begin{tabular}{|c|c|c|c|c|c|c|c|l|}
		\hline
		\multirow{5}{*}{\rotatebox[origin=c]{90}{Nodes}} & $i$ & $\Csite_i$ & $\pcc{max}{i}$ & $\qcc{max}{i}$  & $\lcc{min}{i}$ & $\Cshed_i$ & $\Cload_i$ & Useful $\costLoad{i}$ or $\valueSecondStagePeriod{i}$ values for some $(\kcc{}{i},\lcc{}{i})$ inputs\\ \cline{2-9}\cline{2-9}
		& 1 & 300 & 0 & 0 & 0 & 0 & 0 & N/A\\
		& 2 & 0 & 0.9 & 0.3 & $\sfrac{1}{3}$ & 1000 & 450 & $\costLoad{2}(1,0)=1000,\valueSecondStagePeriod{2}(0,\sfrac{2}{3}) = 850, \valueSecondStagePeriod{2}(0,\sfrac{1}{3}) = 700$\\
		& 3 & 0 & 0.6 & 0.2 & $\sfrac{1}{2}$ & 900 & 300 & $\costLoad{3}(1,0) = 900, \valueSecondStagePeriod{3}(0,\sfrac{1}{2}) = 750$\\
		& 4 & 300 & 0.3 & 0.1 & 1 & 650 & 0 & $\costLoad{4}(1,0) = 650$\\
		\hline
	\end{tabular}}
\resizebox*{0.45\textwidth}{1.4cm}{\begin{tabular}{|c|c|c|c|c|c|c|c|}
		\hline
		\multirow{5}{*}{\rotatebox[origin=c]{90}{Edges}} & $e$ & $\resistance{e}$ & $\reactance{e}$ & \multirow{5}{*}{\rotatebox[origin=c]{90}{DERs}} & $\pgc{max}{\der}$ & $\qgc{max}{\der}$& $\nuc{ref}{\der}$ \\ 
		\cline{2-4}\cline{2-4}\cline{6-8}\cline{6-8}
		& $\{0,1\}$ & 0.1 & 0.2 & & \multirow{2}{*}{0.6} & \multirow{2}{*}{0.2}& \multirow{2}{*}{1.05} \\
		& $\{1,2\}$ & 0.1 & 0.2 &  & & & \\
		& $\{1,3\}$ & 0.1 & 0.2 & & \multirow{2}{*}{0.3} & \multirow{2}{*}{0.1}& \multirow{2}{*}{1.05} \\
		& $\{1,4\}$ & 0.1 & 0.2 & & & & \\
		\hline
\end{tabular}}\resizebox*{0.55\textwidth}{1.4cm}{\begin{tabular}{|c|c|c|c|c|c|c|}
\hline
\multirow{4}{*}{\rotatebox[origin=c]{90}{Scenarios}} & $\scenario$ & $\probFails(\scenario)$ & $\scenario_{\{0,1\}}$ & $\scenario_{\{1,2\}}$ & $\scenario_{\{1,3\}}$ & $\scenario_{\{1,4\}}$ \\ \cline{2-7} \cline{2-7}
& S1 & $\sfrac{1}{3}$ & 1 & 0 & 1 & 1\\
& S2 & $\sfrac{1}{3}$ & 0 & 1 & 1 & 1\\
& S3 & $\sfrac{1}{3}$ & 1 & 1 & 1 & 1\\
\hline
\end{tabular}}
\caption{Parameters of the example 4-node network.}\label{tab:exampleParameters}
\end{table}
Consider a 4-node DN (\Cref{subfig:nominal}) connected to substation node 0.
The voltage bounds for each node $i$ are $\nuc{min}{i} = 0.95$ and $\nuc{max}{i} = 1.05$ (see parameters in \cref{tab:exampleParameters}).
The considered failure scenarios are shown in \Cref{subfig:scenarios} and three of 16 feasible allocations in \Cref{subfig:allocations}, where the DER with larger (resp. smaller) capacity is shown in green (resp. blue).

First, we argue that the optimal allocation without considering line repairs, as is the case in \cite{Lamadrid,Chen}, is to allocate DERs to nodes 2 and 3 (allocation A2). Based on the costs of load shedding, the operator's load preference is in the order $2 \succ 3 \succ 4$. Even if the power consumed by each load is adjusted to be identical at $0.3+0.1\j$ by exercising load control ($\lcc{}{2}=1/3,\lcc{}{3}=1/2,\lcc{}{4}=1$), the value in operating loads 2, 3 and 4 is 700, 750, and 650, respectively; see~\cref{tab:exampleParameters}. If the operator were forced to shed one of three loads, then the operator would be best off shedding load at node 4. Thus without considering repairs, the optimal allocation is to allocate DERs at nodes 2 and 3, and it does not matter which DER is allocated to which node between 2 and 3. However, we show  that this allocation is suboptimal.

\begin{table}[htbp!]
	\resizebox{0.8\textwidth}{!}{\begin{tabular}{|p{0.6cm}|c|c|c|c|c|c|c|c|}
			\hline
			\multirow{2}*{$\first$} &
			\multirow{2}*{$\costFirstStage(\first)$} & \multirow{2}*{$\scenario$} &  \multicolumn{4}{c|}{Stage II cost in period $\period$} & \multirow{2}*{$\costSecondStage(\first,\scenario)$} &  \multirow{2}*{$\soo{}{}(\first)$} \\ \cline{4-7}
			& & & $\costSecondStagePeriod{0}$& $\costSecondStagePeriod{1}$ & $\costSecondStagePeriod{2}$ & $\costSecondStagePeriod{3}$ & & \\
			\hline
			\multirow{3}*{A1} & \multirow{3}*{200} & S1 & 1000 & 450 & 450 & 450 & 2350 & \multirow{3}*{3050} \\
			&  & S2 & 1050 & 450 & 450 & 450 & 2400 & \\
			&  & S3 & 1900 & 1000 & 450 & 450 & 3800 & \\
			\hline
			\multirow{3}*{A3} & \multirow{3}*{100} & S1 & 950 & 450 & 450 & 450 & 2300 & \multirow{3}*{3233} \\
			&  & S2 & 1550 & 950 & 450 & 450 & 3400 & \\
			&  & S3 & 1850 & 950 & 450 & 450 & 3700 & \\
			\hline
	\end{tabular}}\caption{Costs in different periods for allocations A1 and A3, and scenarios S1, S2, and S3. The costs under A2 is a constant of 950 for each period, and each scenario, thereby resulting in total expected loss of 3800.}\label{tab:exampleResults}
\end{table} 

Second, we show how the power flow constraints influence DER allocation. If for some line $e$, $\Pc{}{e}=0.3, \Qc{}{e}=0.1$, then $2(\resistance{e}\Pc{}{e}+\reactance{e}\Qc{}{e}) = 0.1$, i.e., the voltage drop along that line equals 0.1. This constrains the amount of power that can flow along any line. If no DER is allocated to node 1,  then the three loads cannot be simultaneously energized even after load control because of voltage bound violations. For e.g., if $\ysc{}{2} =\ysc{}{3} = 1$, $\ygc{}{21} =1$, and $\ygc{}{32}=1$ (such an allocation may be considered since there is no cost for developing sites at nodes 2 and 3), then for all three loads to be energized, power from the larger DER at node 2 must travel to node 4. This would result in a voltage drop of 0.2 between nodes 2 and 4, and a voltage bound violation. Thus, the larger of the two DERs, i.e. DER 1, should be allocated at node 1 for all three loads to be energized. Hence, under allocation A2, load 4 cannot be re-energized in any scenario. 

Third, we show how the uncertainty in scenarios influence the DER allocation. The summary of the operator costs in various stages under considered allocation strategies and scenarios is shown in \cref{tab:exampleResults}. Note that A1 has the lowest total expected cost, i.e. A1 is the optimal strategy. Also, under A1 the smaller DER would be allocated to node 4. This is somewhat counterintuitive in the sense that the DERs are allocated to costly nodes (i.e. larger $\Csite_i$ values), and in case of node 1, allocated to a node without a load. This can be understood by noting that the line $\{1,4\}$ fails in all scenarios. On the other hand, under allocation A3, the load at node 4 will have to be shed for two time periods in first two scenarios, and for three time periods in the third scenario. Hence, A3 is clearly a suboptimal allocation. 

	\section{Solution Approach for \eqref{eq:sooFirstStage}} \label{sec:solutionApproach}
	

In this section, we outline our approach for obtaining solutions to the two-stage program \eqref{eq:sooFirstStage}. We use the sample average approximation (SAA) method to solve \eqref{eq:sooFirstStage} \cite{Shapiro}, which takes a random subset of the scenario set $\setScenarios$ as input. Specifically, the SAA problem is given by:
\begin{align}\label{eq:sooEmpirical}\tag{SAA}
\begin{aligned}
\mmin_{\first\in\First} \ \{
\saaSoo{\nscenario}{}(\first):= \costFirstStage(\first) + (1/\nscenario)\ssum_{\scenarioIdx \in \setScenariosSmall}  \costSecondStage\left(\first, \scenario\right) \}.
\end{aligned}
\end{align}
Here, $\setScenariosSmall \subset \setScenarios$ is a random subset of the set of failure scenarios, $\nscenario \coloneqq |\setScenariosSmall|$, and $\saaSoo{\nscenario}{}(\first)$ is the SAA objective. 

We solve \eqref{eq:sooEmpirical} rather than obtain an exact solution to \eqref{eq:sooFirstStage}, because calculation of $\mathbb{E}_{\scenario\sim\probFails}[\costSecondStage(\first, \scenario)]$ in \eqref{eq:sooFirstStage} is computationally intractable for  large networks. The number of Stage II problems to solve under allocation $\first$ is $2^\NN$. If $\nscenario \ll |\setScenarios|$, \eqref{eq:sooEmpirical} requires much less computation time to solve than \eqref{eq:sooFirstStage}. A naive approach to solve \eqref{eq:sooEmpirical} is to solve \eqref{eq:sooSecondStage} for each scenario $\scenarioIdx\in\setScenariosSmall$ under each strategy $\first\in\First$. This is inefficient because the number of feasible solutions $|\First|$ increases exponentially with number of DERs. 


\begin{figure}[htbp!]
	\centering
    \tikzstyle{decision} = [diamond, draw, fill=white!20, 
        text width=4.5em, text badly centered, node distance=3cm, inner sep=0pt]
    \tikzstyle{block} = [rectangle, draw, fill=white!20, 
        text width=10em, text centered, rounded corners, minimum height=2em]
    \tikzstyle{exit} = [rectangle, fill=white!20, 
        text width=1em, text centered, rounded corners, minimum height=4em]
    \tikzstyle{line} = [draw, -latex']
    \tikzstyle{input} = [rectangle, dashed, draw, fill=white!20, 
        text width=8em, text centered, rounded corners, minimum height=4em]
\resizebox*{\textwidth}{!}{
    \begin{tikzpicture}[node distance = 1.5cm, auto]
    \small
        \node [input] (input) {Resource budget, crew capacity};
        \node [input, left of=input, text width=8em, node distance = 3.5cm] (input2) {Network topology, parameters, costs};
        \node [input, right of=input, text width=8em, node distance = 3.5cm] (input3) {Failure scenarios};
        \node [block, below of=input, node distance = 1.5cm] (MP) {Solve \eqref{eq:masterProblem} w/ Benders cuts};
        \node [decision, below of=MP, node distance = 1.5cm, aspect = 2] (evaluate) {$\First^\star_\iter$ = $\varnothing$};
        \node [block, below of=evaluate, node distance = 1.5cm] (GS-MIP) {Solve \eqref{eq:sooSecondStage} w/ greedy approach};
        \node [block, below of=GS-MIP] (S-MIP) {Solve \eqref{eq:sooSecondStage} w/ upper bounds};
        \node [block, left of=S-MIP,node distance = 5cm] (S-LP) {Solve \eqref{eq:sooSecondStageLP}};
        \node [block, above of=S-LP] (updateUB) {Update current solution $\tilde{\soo{}{}}$};
        \node [block, left of=evaluate, node distance = 5cm] (addBD) {Add Benders cut};
        \node [exit, right of=evaluate, node distance = 2cm] (exit1) {exit};
        \path [line] (input) -- (MP);
        \path [line] (input2) -- (MP);
        \path [line] (input3) -- (MP);
        \path [line] (MP) -- node [left] {$\First^\star_\iter$}(evaluate);
        \path [line] (evaluate) -- node [right]{$\first^\star_\iter\in\First^\star_\iter$}(GS-MIP);
        \path [line] (evaluate) -- node [left]{no}(GS-MIP);
        \path [line] (GS-MIP) -- node [right] {$\recourseUb(\first^\star_\iter, \scenarioIdx)$, $\forall\scenarioIdx$}(S-MIP);
        \path [line] (S-MIP) -- node [above] {$\yc{}{}^\star$, $\xc{}{\disc}^\star$}(S-LP);
        \path [line] (S-LP) -- node [left] {Benders cut}(updateUB);
        \path [line] (updateUB) -- node [near start] {}(addBD);
        \path [line] (evaluate) -- node {yes}(exit1);
         \path [line] (addBD) |- node [] {}(MP);
  	 \end{tikzpicture}
	}
    	\caption{Overview of the proposed solution approach.}
	\label{fig:algFlow}
\end{figure}
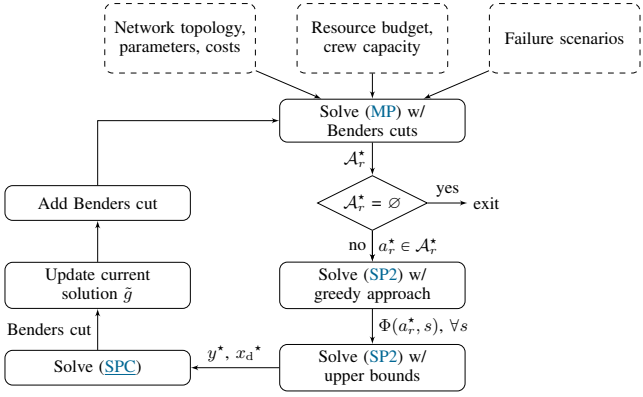

We propose a more efficient approach based on L-shaped Benders decomposition (LBD), which can output the optimal solution to \eqref{eq:sooEmpirical} by potentially considering a smaller number of Stage I strategies. Akin to LBD for two-stage stochastic programs, our approach alternates between a master problem and sub-problems at each iteration $\iter$ (see~\Cref{fig:algFlow}). The master problem is defined as:
	\begin{equation}\label{eq:masterProblem}\tag{MP}
		\textstyle\First_\iter^\star = \argmin_{\first\in\First}\ \costFirstStage(\first) 
		\quad \text{ s.t. Benders cuts},
	\end{equation}	
	where the Benders cuts are defined in \Cref{sec:benders}. At the start of the first iteration, there are no Benders cuts and the current solution $\tilde{\soo{}{}}$ to \eqref{eq:sooEmpirical} is initialized to $\infty$. If $\First_\iter^\star$ is not empty, an optimal solution $\first^\star_\iter\in\First_\iter^\star$ is used as input to solve the sub-problem \eqref{eq:sooSecondStage} for each scenario $\scenario\in\setScenarios'$. Then, we solve a linear program \eqref{eq:sooSecondStageLP} $\forall\scenario\in\setScenarios'$, formed using \eqref{eq:sooSecondStage} with the discrete variables fixed. We update the solution $\tilde{\soo{}{}}$ as: $\tilde{\soo{}{}}\leftarrow\min\{\tilde{\soo{}{}}, \saaSoo{\nscenario}{}(\first_\iter^\star)\}$ where $\saaSoo{\nscenario}{}(\first_\iter^\star)$ denotes the objective at iteration $\iter$. A new Benders cut is added to \eqref{eq:masterProblem} using the solutions to \eqref{eq:sooSecondStage} and \eqref{eq:sooSecondStageLP}. The algorithm terminates when the Benders cuts renders all first-stage solutions infeasible (i.e., $\First^\star_\iter = \emptyset$). If termination occurs at iteration $\niter$, we define the set of solutions from all iterations as $\First_\niter=\{\first^\star_1, \cdots, \first^\star_{\niter-1}\}$.
The solution from $\First_\niter$ that produces the smallest objective is the optimal solution. 

	The greatest computational burden in LBD arises from solving \eqref{eq:sooSecondStage} for each scenario. In \Cref{sec:algorithms}, we discuss a greedy approach to decrease computation required for \eqref{eq:sooSecondStage}. 
	
	\subsection{Formation of Benders cuts}\label{sec:benders}	
Given an allocation $\first^\star$ and scenario $\scenarioIdx$, the optimal objective for a sub-problem \eqref{eq:sooSecondStage} is $\costSecondStage\left(\first^\star, \scenario\right)$. The corresponding optimal solution is given by $(\xc{\scenarioIdx}{})^\star$ and $(\yc{\scenarioIdx}{})^\star$, where  $(\xc{\scenarioIdx}{})^\star$ can be partitioned into $(\xc{\scenarioIdx}{\disc})^\star$ and $(\xc{\scenarioIdx}{\cont})^\star$ to denote the discrete (resp. continuous) variables. The optimal objective of \eqref{eq:sooSecondStage} can be rewritten as $\ssum_{\period\in\setPeriods}
(\costSecondStageLin{\top}{xd,k}(\xc{\ts}{\disc})^\star + \costSecondStageLin{\top}{xc,k}(\xc{\ts}{\cont})^\star)$.
After solving \eqref{eq:sooSecondStage}, we solve the Stage II problems with the discrete variables $(\yc{\scenarioIdx}{})^\star$ and $(\xc{\scenarioIdx}{\disc})^\star$ fixed. The resultant LP is:
\begin{align}\label{eq:sooSecondStageLP}\tag{\underline{SPC}}
\begin{aligned}
\mmin \ & \ssum_{\period\in\setPeriods} \costSecondStageLin{\top}{xc,k}\xc{\ts}{\cont} \\ \text{s.t.} \ & \secondStageConstraints{}{xc}\xc{\scenario}{\cont}   \ge \ \secondStageRh - \secondStageRhA\first^\star - \secondStageConstraints{}{y}(\yc{\scenario}{})^\star - \secondStageConstraints{}{xd}(\xc{\scenario}{\disc})^\star, \\
\end{aligned}
\end{align}	
where $\secondStageConstraints{}{xd}$ and $\secondStageConstraints{}{xc}$ are the columns of $\secondStageConstraints{}{x}$ corresponding to the discrete and continuous variables, respectively. 

A Benders cut is formed using the discrete variables from \eqref{eq:sooSecondStage} and the dual solution from \eqref{eq:sooSecondStageLP}:
\begin{equation}
\begin{aligned}
\hspace{-0.3cm}\costFirstStageCoeff^\top\first+
\ssum_{\scenarioIdx\in\setScenarios}[\costSecondStageLin{\top}{xd,k}(\xc{\ts}{\disc})^\star +\transpose{\tilde{\secondStageRh}(\first,\scenario)}(\lambda^\scenarioIdx)^\star]\leq \lossLP-\epsilon,
\end{aligned}
\end{equation}
where $\tilde{\secondStageRh}(\first,\scenario) = \secondStageRh - \secondStageRhA\first - \secondStageConstraints{}{y}(\yc{\scenario}{})^\star - \secondStageConstraints{}{xd}(\xc{\scenario}{\disc})^\star$ is the right-hand side of \eqref{eq:sooSecondStageLP}, $(\lambda^\scenarioIdx)^\star$ is the dual solution to \eqref{eq:sooSecondStageLP}, loss $\lossLP = \costFirstStageCoeff^\top\first^\star+ \ssum_{\scenario\in\setScenariosSmall}\costSecondStage\left(\first^\star, \scenarioIdx\right)$, and $\epsilon\approx10^{-6}$ is a small positive number. This cut renders $\first^\star$ infeasible because the resulting objective value  $\soo{}{}(\first^\star)$ is exactly $\lossLP$. 	

\subsection{Greedy Approach to Stage II problem} \label{sec:algorithms}
We now describe our greedy approach to decrease the computation time required for the Stage II MIP \eqref{eq:sooSecondStage}. The number of binary variables in \eqref{eq:sooSecondStage} increases quadratically with the network size, because it requires at least  $\mathcal{O}(\sfrac{\abs{\E}}{\crewBudget})$ periods for line repairs, and there are $\mathcal{O}(\abs{\N}+\abs{\E})$ binary variables for each period. Thus, off-the-shelf MIP solvers, which typically implement branch-and-bound (B\&B) algorithms, require significant computational time to solve \eqref{eq:sooSecondStage}. Our greedy approach finds a feasible solution to \eqref{eq:sooSecondStage} so that the corresponding objective value can be used as a reasonable upper bound, which can significantly reduce the number of B\&B nodes explored. 

The greedy approach entails sequentially obtaining a period-wise line repair and power dispatch solution from first to last period. We take advantage of the fact that there are no inter-period dependencies in the power flow constraints, which permits us to decompose \eqref{eq:sooSecondStage} into an MIP for each period. At each period $\period$, the solutions $\yc{\period,\scenarioIdx}{}$ and $\xc{\period,\scenarioIdx}{}$ myopically minimize the Stage II cost for the period, i.e. $\costSecondStagePeriod{\period}$. The constraints on $\yc{\period,\scenarioIdx}{}$ are only dependent on the operational states $\klinec{\period-1,\scenario}{}$ from the previous period $\period-1$, and the constraints on $\xc{\period,\scenarioIdx}{}$ only depend on the state $\klinec{\ts}{}$. With a slight abuse of notation, for $\period \in \setPeriods$, we denote by  $\Yc{\period}{}(n)$ a set of vectors $\yc{\ts}{}$ which satisfy only those constraints among \eqref{eq:norepairFirstPeriod}-\eqref{eq:repairBinaryConstraint} that involve variables $\yc{\ts}{}$, where $ \klinec{\period-1,\scenario}{} = n$ is fixed. Similarly, let  $\Xc{\period}{}(\first,\scenarioIdx,\yc{\ts}{})$ denote a set of vectors $\xc{\ts}{}$ which satisfy only those constraints among  \eqref{eq:derSiteActiveContribution}-\eqref{eq:voltageDrop} that  involve variables $\xc{\ts}{}$. 

Now, consider a fixed pair $(\first,\scenario) \in \First\times\setScenarios$. Then, 
for $\period=0,\cdots,\nperiod$, we solve the following MIP:
\begin{align}\label{eq:greedy}
\begin{aligned}\small
& \recourseUb_\period\big(\first, (\klinec{\period-1,\scenario}{})^\star\big) \ := \ \min_{\xc{\period,\scenarioIdx}{},\yc{\period,\scenarioIdx}{}}
\quad\costSecondStageLin{\top}{x,k}\xc{\ts}{}\\
&\text{ s.t. } \ \yc{\period,\scenarioIdx}{}\in \Yc{\period}{}\big( (\klinec{\period-1,\scenario}{})^\star\big), \  \xc{\period,\scenarioIdx}{} \in \Xc{\period}{}\left(\first,\scenarioIdx,\yc{\period,\scenarioIdx}{}\right),
\end{aligned}
\end{align}
where $(\klinec{\period-1,\scenario}{})^\star$ is part of the optimal solution to the MIP $\recourseUb_{\period-1}$ in the previous iteration. For $\period = 0$, we consider $(\klinec{\period-1,\scenario}{})^\star = \scenario$, since there are no repairs before $\period=0$. 

Then, the sum of the optimal solutions to the greedy problem, $\recourseUb(\first, \scenarioIdx) = \recourseUb_0\left(\first, \scenarioIdx\right) + \ssum_{\period = 1}^{\nperiod}  \recourseUb_\period(\first,  (\klinec{\period-1,\scenario}{})^\star)$ is an upper bound to the Stage II objective. Thus, the following cut can be used for the Stage II problems:
\begin{align}\label{eq:SMIP_UB}
\begin{aligned}\small
\ssum_{\period\in\setPeriods}
\costSecondStageLin{\top}{x,k}\xc{\ts}{} \leq \recourseUb(\first, \scenarioIdx).
\end{aligned}
\end{align} 
As shown in \Cref{fig:algFlow}, we first obtain an upper bound to \eqref{eq:sooSecondStage} using the greedy approach, then solve \eqref{eq:sooSecondStage} with the cut \eqref{eq:SMIP_UB} as an added constraint. Then, we use the solution to \eqref{eq:sooSecondStage} in order to solve \eqref{eq:sooSecondStageLP} and obtain a Benders cut (\Cref{sec:benders}).	
%
%
	\section{Computational Study}\label{sec:compStudy}
In this section, we discuss our computational study. In \Cref{sec:sysPerf}, we analyze the system performance attained by our approach, compared to simpler approaches that search the set of feasible allocation solutions less exhaustively. In \Cref{sec:validation}, we evaluate the performance of the greedy approach. For our study, we use a modified IEEE 36-node test feeder.\footnote{12 out of 36 randomly-chosen nodes have one load each. We otherwise use the same parameters as in our previous work \cite{SGC}.}

\subsection{System performance evaluation}\label{sec:sysPerf}
We compare solutions from four approaches: (1) Simple enumeration ($\se$): best solution from the set of  allocations obtained by simple enumeration; (2) solution to LBD with Greedy Approach ($\bwga$), outlined in \Cref{sec:solutionApproach}; (3) `Best of Random Allocations' ($\bora$): best solution from a set of randomly-sampled allocations; and (4) `Single Allocation' ($\sa$): one pre-determined allocation such that the DERs
have an even spatial distribution across the DN.

\vspace{0.1cm}
\begin{figure}[htbp!]
	\centering
	\includegraphics[width=0.6\textwidth]{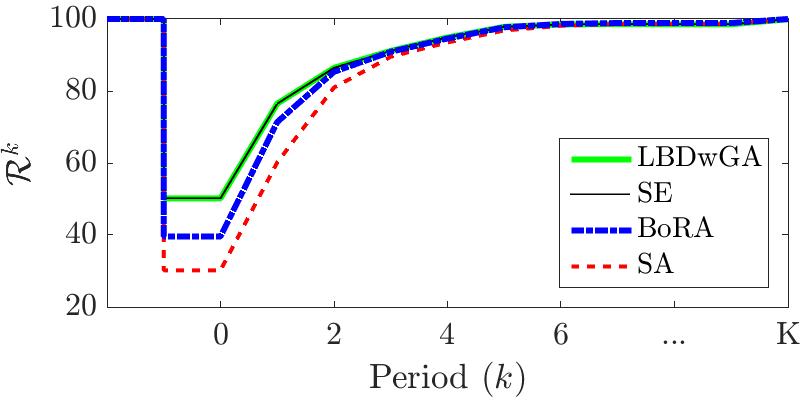}
	\vspace{-0.2cm}
	\small\caption{Average system performance of the 36-node DN under allocations given by $\bwga$, $\bruteforce$, $\bora$, and $\sa$. We use the parameters $\genBudget = 3$,  $\crewBudget = 2$, and $\nscenario = 10$.}
	\label{fig:sysPerf}
\end{figure}

\Cref{fig:sysPerf} demonstrates how system performance $\systemPerformance$ evolves over the set of periods under the four approaches. Before the storm occurs, the network is in nominal operation and $\systemPerformance = 100\%$. After the storm ($\period=0$), $\systemPerformance$ is at a minimum, and improves in subsequent periods with each set of line repairs. Once all the damaged lines are repaired, $\systemPerformance$ is almost (but not fully) restored. Finally, $\systemPerformance$ returns to 100\% following reconnection of the DN to the bulk power grid at $\period = \nperiod$. As expected, $\bwga$ and $\bruteforce$ have equivalent system performance and outperform the other methods.

\subsection{Evaluation of greedy approach}\label{sec:validation} 
\begin{figure}[htbp!]
	\centering
	\vspace{-0.35cm}
	\subfloat{\includegraphics[trim=0cm 0cm 0cm 1cm, clip, width=0.5\textwidth, scale=0.12]{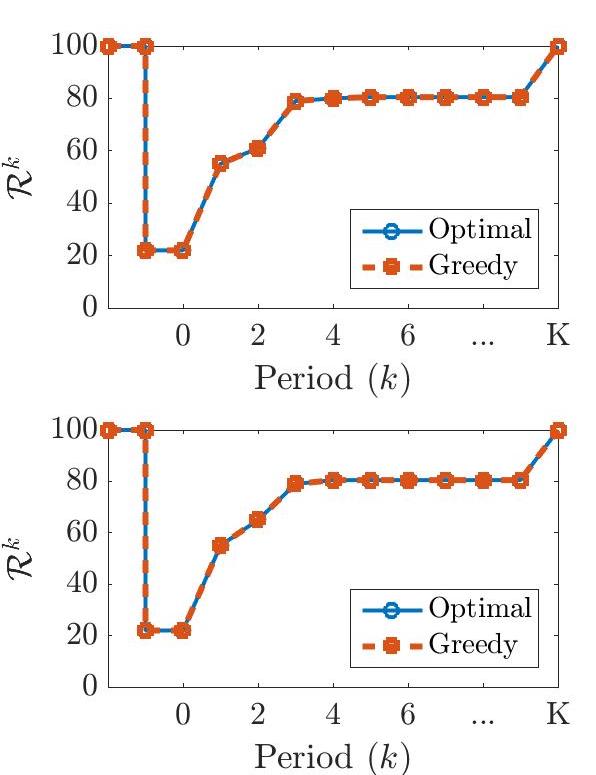}}
	\subfloat{\includegraphics[width=0.5\textwidth, scale=0.12]{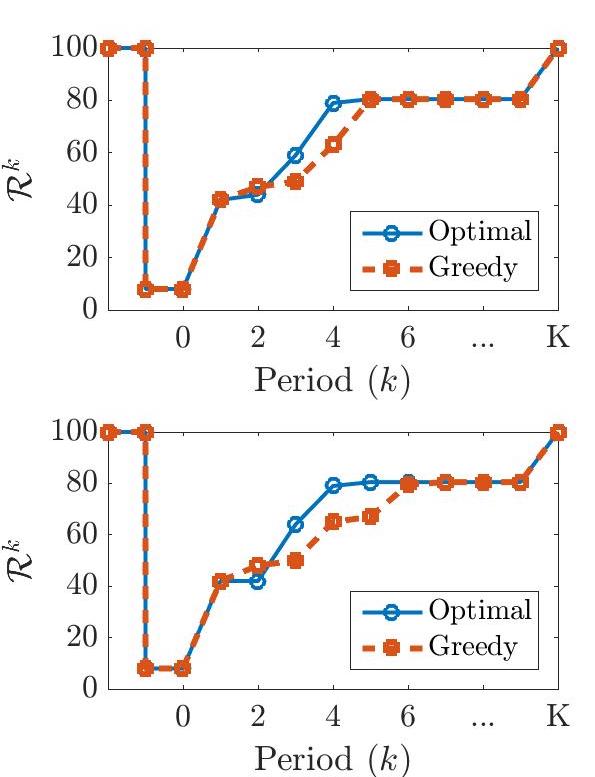}}		
	\setcounter{subfigure}{0}	
	\vspace{-0.6cm}
	\caption{System performance under Scenario 1 (\textit{left column}) and Scenario 2 (\textit{right column}). Parameters are $\genBudget=1,\crewBudget=2$ (\textit{top row}) or $\genBudget=1,\crewBudget=3$ (\textit{bottom row}). }
	\label{fig:greedyComp}
\end{figure}

The greedy solution provides an upper bound to the optimal value of \eqref{eq:sooSecondStage}. In \Cref{fig:greedyComp}, we compare the system performance of the optimal and greedy solutions under two different scenarios. For Scenario 1, the greedy solution exactly matches the optimal solution. For Scenario 2, the greedy solution is suboptimal, but the difference between the greedy and optimal solutions' system performance is small. Although the greedy solution outperforms the optimal solution at $\period=2$ (see right-hand plots in \cref{fig:greedyComp}), the total expected cost will be higher in the former case. 

A smaller difference between the system performance of the optimal and greedy solutions indicates a tighter upper bound. In Scenario 1, the greedy upper bound ensures that the greedy solution is the only feasible solution, because the greedy and optimal solutions are the same. In Scenario 2, the greedy upper bound renders a large number of repair schedules infeasible, but multiple feasible schedules remain.

	\section{Concluding remarks}\label{sec:concludingRemarks}
	This work presents a solution approach to the integrated pre-storm resource allocation and post-storm repair and dispatch problem for improving resilience of electricity distribution networks against storms. The problem is formulated as a two-stage stochastic multi-period program. 
	The solution approach involves Sample Average Approximation (SAA), L-shaped Benders decomposition, and a greedy approach to reduce the cost of solving Stage II recourse subproblems. 
	
	We plan to extend our work in three directions, to permit scalability of our approach to larger networks. First, we will provide provable guarantees on the upper bound formed by the greedy approach, and discuss how the greedy solutions may be improved. This would decrease the computation time required for scenario-wise sub-problems. Second, we will focus on decreasing the number of iterations in Benders decomposition. In particular, we will form lower bounds to the two-stage MILP objective, which may permit us to achieve convergence before our current approach exhausts all feasible first-stage allocations. Finally, we will evaluate accuracy of SAA solutions using optimality gap estimates.
	\bibliography{IEEEabbrv}

\end{document}